\documentclass[singlecolumn,11pt,aps,prd,superscriptaddress,preprintnumbers,amsmath,amssymb,nofootinbib]{revtex4-2}

\usepackage{epsfig}  
\usepackage{graphicx}
\usepackage{color}
\usepackage[dvipsnames]{xcolor}
\usepackage{float}
\usepackage{amsfonts}
\usepackage{amsmath}
\usepackage{mathrsfs}
\usepackage{slashed}
\usepackage{soul}
\usepackage{braket}
\usepackage{dsfont}
\usepackage{hyperref}

\begin{document}

\title{Quantifying Imaginarity in Neutrino Systems}

\author{Ashutosh Kumar Alok}
\thanks{Ashutosh Kumar Alok passed away on 10 October 2024.}
\affiliation{Indian Institute of Technology Jodhpur, Jodhpur 342037, India}

\author{Trambak Jyoti Chall}
\email{chall.1@iitj.ac.in}
\affiliation{Indian Institute of Technology Jodhpur, Jodhpur 342037, India}

\author{Neetu Raj Singh Chundawat}
\thanks{Corresponding Author}
\email{chundawat.1@iitj.ac.in}
\affiliation{Indian Institute of Technology Jodhpur, Jodhpur 342037, India}
\affiliation{Institute of High Energy Physics, Chinese Academy of Sciences, Beijing 100049, China}
\affiliation{Kaiping Neutrino Research Center, Guangdong 529386, China}

\author{Yu-Feng Li}
\email{liyufeng@ihep.ac.cn}
\affiliation{Institute of High Energy Physics, Chinese Academy of Sciences, Beijing 100049, China}\affiliation{School of Physical Sciences, University of Chinese Academy of Sciences, Beijing 100049, China}

\begin{abstract}
It is a fundamental question why quantum mechanics employs complex numbers rather than solely real numbers. In this work, we conduct the {\it first analysis} of imaginarity quantification in neutrino flavor and spin-flavor oscillations. As quantum systems in coherent superposition, neutrinos are ideal candidates for quantifying imaginarity within the resource theoretic framework, using measures such as the $\ell_1$-norm and the relative entropy of imaginarity. We show that in the case of two-flavor mixing, these measures of imaginarity are nonzero. 
The measures of imaginarity reach their extreme values when the probabilistic features of quantum theory are fully maximized, i.e., both the transitional and survival probabilities are approximately equal. Our study reveals that the imaginarity, as a resource, can be harnessed not solely from the presence of a complex phase in the mixing matrix but also from the intrinsic quantum dynamics of time evolution itself. We further extend our analysis to explore the dynamics of three-flavor neutrino mixing, incorporating the effects of a nonzero $CP$ phase. 

\end{abstract}

\maketitle
\newpage



\section{Introduction}
The complex numbers are mathematical representations extensively used in various fields of physics. In classical physics, imaginary numbers have been employed to simplify models of oscillatory motion and wave dynamics. In quantum mechanics, wave-particle duality is a fundamental aspect necessitating the use of complex numbers to describe the states, dynamics, and interactions of quantum systems. The postulates of quantum mechanics state that a quantum system is described by a complex Hilbert space. Since its inception, there has been an ongoing debate about whether complex numbers are used merely for mathematical convenience or if they are essential for describing quantum systems. In other words, are complex numbers indispensable, or is it possible to develop a consistent formulation using only real numbers? Essentially, can quantum physics be reformulated using only real numbers?
 
It is, therefore, essential to thoroughly investigate the role of complex numbers in quantum systems. This analysis is vital for comprehending the fundamental aspects of a quantum state, such as superposition and entanglement, and also for the practical application of quantum information science in quantum technologies. For example, quantum computers can solve certain problems much faster than classical computers, and a quantitative analysis of imaginarity in these systems could provide new insights into the quantum features necessary for achieving quantum speedup~\cite{Wu:2021jjy}.

In recent years, quantum information theory has increasingly focused on treating uniquely quantum phenomena such as entanglement and nonlocality as operational resources that enable tasks unattainable within classical physics, for example quantum teleportation. Quantum resource theories (QRTs) provide a systematic framework to analyze and quantify these resources~\cite{Chitambar:2018rnj, Gour:2024cbn}. Within this approach, one can determine the amount of a given resource contained in a quantum state, identify optimal procedures for resource distillation, and study the possible interconversion of resource states under a set of allowed free operations. Significant effort has therefore been devoted to developing methods for quantifying various quantum resources, including entanglement, coherence, thermodynamic resources, and Bell nonlocality within the framework of QRT.
 
 Given the unique significance of imaginary numbers in quantum theory, recent developments have led to a comprehensive formulation of imaginarity in quantum physics and quantum information theory, see, for e.g., ~\cite{Hickey,Wu,Xu:2023hek,Xue:2021zmd,Zhu:2020iml,Li:2021uof,Renou:2021dvp,Chen:2022rco,Kondra,Xu:2023xdb,Wu:2023lyi,Xu:2024yqx,Zhang:2024maj}. The QRT of imaginarity was developed in~\cite{Hickey}, where the framework is set to quantify and control particular properties of a quantum mechanical system and identify the information processing tasks achievable by utilizing these properties. This QRT of imaginarity stems from the imaginary components present in the density matrix of a quantum state. The imaginarity theory is intrinsically connected to the coherence theory because the imaginary parts of the density matrix are always found in the off-diagonal elements. By examining only these imaginary off-diagonal elements, we can discern how the relative phase between measurement basis states impacts the system's potential dynamics.

In~\cite{Wu:2021jjy}, using the robust framework of QRT, it was shown that, under reasonable conditions, quantum states composed entirely of real elements are easier to create and manipulate, thereby providing operational significance to the resource theory of imaginarity. As an application, they illustrated that imaginarity is crucial in state discrimination. They showed that there are real quantum states that can be perfectly distinguished through local operations and classical communication but cannot be differentiated with any nonzero probability if one of the parties lacks access to imaginarity. This was experimentally validated with linear optics, where they distinguished various two-photon quantum states using local projective measurements. Their findings confirmed that complex numbers are essential in quantum mechanics.

Neutrinos as quantum systems provide a natural platform to explore nonclassical features~\cite{Gangopadhyay:2013aha,Formaggio:2016cuh,  Blasone:2021mbc, Blasone:2022iwf, Naikoo:2019eec, Ming:2020nyc, Alam:2026bxn}, in this work, we present the first analysis of imaginarity quantification within neutrino systems using resource theoretic approach. We study the imaginarity theory for neutrino states undergoing flavor oscillations (FOs). The complex numbers naturally arise in the time evolution of neutrino states through phase factors, so it is reasonable to test the measures of imaginarity in the two-flavor approximation as well. We use two metrics of imaginarity: the $\ell_1$ norm of imaginarity, which captures the imaginary components of the density matrix elements, and the relative entropy of imaginarity, which quantifies the difference in the von Neumann entropy
between the real component of a quantum state's density matrix and its complete density matrix. 

Meanwhile, neutrinos with nonzero magnetic moments may undergo spin-flavor oscillations (SFOs) in an external magnetic field. We examine the role of imaginarity in neutrino systems with additional degrees of freedom in the context of SFOs. We further delve into the three-flavor regime and analyze the imaginarity in the system with a non-zero complex $CP$-violating phase.

The structure of this work is as follows. Section II introduces the imaginarity measures employed in our analysis. In Section III, we present a detailed description of neutrino oscillations in vacuum, as well as spin-flavour oscillations of neutrinos with magnetic moments propagating through a magnetic field. Section IV explores and interprets the behaviour of the imaginarity measures across various neutrino systems within both two- and three-flavour oscillation frameworks. Finally, Section V offers our concluding remarks.


\section{Imaginarity Theory}
We first introduce all the imaginarity quantifiers employed in this work. In the imaginarity theory, a fixed orthonormal basis \(\{ |m\rangle \}_{m=1}^{d}\) is introduced for a \(d\)-dimensional complex Hilbert space \(\mathcal{H}\). In this basis, a quantum state is represented by $\rho = \sum_{m,n=1}^{d} \langle m|\rho|n \rangle |m\rangle \langle n|$. 
Given the Hermitian relation of \(\rho = \rho^{\dagger}\), we can obtain that \((\text{Re} \, \rho)^{T} = \text{Re} \, \rho\) and \((\text{Im} \, \rho)^{T} = -\text{Im} \, \rho\).
A quantum operation \(\phi\) is defined by the set of Kraus operators \(\{K_l\}_l\) satisfying the condition of \(\sum_l K_l^{\dagger} K_l \leq I\), where \(I\) is the identity operator. The quantum operation representing a completely positive and trace-preserving map \(\phi\) is referred to as a quantum channel when \(\sum_l K_l^{\dagger} K_l = I\), which ensures the conservation of the total probability.
Note that the quantum operation \(\phi\), transforms \(\rho\) into a new state \(\phi(\rho) = \sum_l K_l \rho K_l^{\dagger}\) that is not necessarily normalized~\cite{Hickey}. 

In the context of imaginarity theory, a free operation is a ``real" quantum operation that cannot generate imaginarity in quantum states that are initially real~\cite{Wu:2021jjy}.
This means that if a quantum state \(\rho\) is real, consisting solely of real elements in the specified basis, then applying a real operation \(\phi\) will preserve its reality, ensuring that no imaginary components are introduced. These operations neither consume nor generate resources within the system. Therefore, quantifying the extent to which a state deviates from being ``free" indicates how resourceful that state is. 
An imaginarity measure is a quantifier used to evaluate the role of complex numbers in quantum theory. For an imaginarity measure, denoted as \(\mathcal{I}\), to serve as a meaningful and consistent tool, it must satisfy several criteria listed below:

\begin{itemize}
    \item \textbf{Faithfulness:}
    The measure of imaginarity should be non-negative for any quantum state \(\rho\), i.e.,
    \begin{equation}
    \mathcal{I}(\rho) \geq 0.
   \end{equation}
    The equality holds only for a real state, implying that the measure is zero if and only if \(\rho\) is a real state.
    This condition guarantees that the measure effectively indicates the presence of imaginary components in the state.
    
    \item \textbf{Imaginarity Monotonicity:} 
    The measure of imaginarity should not increase under the application of real operations.
    Formally, if \(\phi\) is a real channel, then \begin{equation}
    \mathcal{I}[\phi(\rho)] \leq \mathcal{I}(\rho).
    \end{equation} 
    This condition ensures that when a real channel (comprising real Kraus operators) is applied to a quantum state, the imaginarity measure does not increase.

\end{itemize}
Additionally, the following properties can also be investigated to further understand and characterize the measure of imaginarity:

\begin{itemize}
    
    \item \textbf{Strong Imaginarity Monotonicity:} 
    This property requires that the measure of imaginarity for a quantum state \(\rho\) should be greater than or equal to the weighted sum of the imaginarity measures of the resultant states after applying real Kraus operators.
    Mathematically, it is expressed as 
    \begin{equation}
    \mathcal{I}(\rho) \geq \sum_j p_j \mathcal{I}(\rho_j),
    \end{equation}
    where \(p_j = \text{Tr}[K_j \rho K_j^{\dagger}]\) and \(\rho_j = K_j \rho K_j^{\dagger} / p_j\) with the real Kraus operators \(K_j\).
    This ensures that the measure does not increase under probabilistic mixtures of real operations, reflecting the consistent reduction of imaginarity.

    \item \textbf{Convexity:}
    This property ensures that mixing quantum states should not increase the measure of imaginarity. Formally, it is written as
    \begin{equation}
    \mathcal{I}\left(\sum_j p_j \rho_j\right) \leq \sum_j p_j \mathcal{I}(\rho_j).
    \end{equation}
    This reflects the concept that the imaginarity of a convex combination of states (a mixture) should not exceed the weighted sum of the imaginarities of the individual states, ensuring that imaginarity is a resource that cannot be ``created'' by mixing states.

    \item \textbf{Additivity:}
    This property implies that the measure of imaginarity should be additive for independent quantum systems. For two quantum states \(\rho_1\) and \(\rho_2\), the measure of their combined state should be the sum of their individual measures, weighted by the probabilities of the states. It is expressed as:
    \begin{equation}
      \mathcal{I}(p \rho_1 \oplus (1 - p) \rho_2) = p \mathcal{I}(\rho_1) + (1 - p) \mathcal{I}(\rho_2).
    \end{equation}
    This property is crucial for understanding how imaginarity scales with the size of the quantum system and ensures consistent behavior when dealing with larger composite systems.
\end{itemize}

For real operations which adhere to completely positive trace-preserving (CPTP) quantum operations, it can be shown that the $\ell_1$-norm of imaginarity can be expressed as:
\begin{equation}
\mathcal{I}_{\ell_1}(\rho) = \sum_{i \neq j} |\text{Im}(\rho_{ij})|\,.
\end{equation}
Here, \(\text{Im}(\rho_{ij})\) represents the imaginary part of the density matrix element \(\rho_{ij}\). This expression encapsulates the essence of imaginarity by summing the absolute values of the imaginary components of the off-diagonal elements in the density matrix. On the other hand, the relative entropy of imaginarity measures the entropic distance between the quantum state and its real counterpart, which can be expressed as:
\begin{equation}
\mathcal{I}_r(\rho) = S(\text{Re}(\rho)) - S(\rho), 
\end{equation}
where \(S(\rho) = -\mathrm{Tr}[\rho \log \rho]\) is defined as the von Neumann entropy~\cite{Xue:2021zmd}, and 
 \( \text{Re}(\rho) \) is the real part of \( \rho \), which is expressed as 
\(\text{Re}(\rho)=(\rho + \rho^{\mathsf{T}})/{2}. \)

For a pure quantum state $\ket{\psi}$, the entropic measure reduces to a $|\langle \psi^* | \psi \rangle|$-dependent form given by
\begin{equation}
\mathcal{I}_r(|\psi\rangle) 
= 1 - \frac{1 + |\langle \psi^* | \psi \rangle|}{2} 
   \log \left( 1 + |\langle \psi^* | \psi \rangle| \right) \nonumber  - \frac{1 - |\langle \psi^* | \psi \rangle|}{2} 
   \log \left( 1 - |\langle \psi^* | \psi \rangle| \right) \,.
\label{rel}
\end{equation}

Both of these observables \(\mathcal{I}_{\ell_1}(\rho)\) and \(\mathcal{I}_r(\rho)\) will be nonzero only if the quantum state \(\rho\) contains imaginary components, thereby serving as effective metrics of the quantum state's imaginarity. In the following sections, we shall establish the properties of imaginarity in the neutrino systems of FOs and SFOs.

\section{Neutrino Flavor and Spin Flavour Oscillations}
The neutrino flavor state \(|\nu_{\alpha}^{h}(l)\rangle\) is described by a coherent superposition of three mass eigenstates \(|\nu_{i}^{h}(l)\rangle \) as~\cite{ParticleDataGroup:2024cfk}
\begin{equation}
    |\nu_{\alpha}^{h}(l)\rangle=\sum_{i=1}^3 U^{*}_{\alpha i}|\nu_{i}^{h}(l)\rangle\,,
    \label{3mix}
\end{equation}

where $\alpha=(e,\mu,\tau)$ is the flavor index, $h$ is the helicity index of the neutrino state, and $l$ is the spatial variable that describes the distance from the neutrino production point. $U_{\alpha i}$ is a $3 \times 3$ unitary matrix representing the lepton flavor mixing, commonly referred to as the Pontecorvo–Maki–Nakagawa–Sakata (PMNS) matrix~\cite{Maki:1962mu,Pontecorvo:1967fh}. It can be parametrized with four physics quantities, three mixing angles $\theta_{12},\theta_{13},\theta_{23}$ and the complex $CP$ violating phase $\delta_{\rm CP}$. The coherent spatial evolution of neutrino flavor states leads to the phenomenon of neutrino FOs, which has been observed in various neutrino oscillation experiments~\cite{ParticleDataGroup:2024cfk}.

The observation of FOs implies that neutrinos have masses and may result in the generation of a small but finite magnetic dipole moment for neutrinos through high-order quantum radiative corrections~\cite{Giunti:2014ixa,Giunti:2015gga,Giunti:2024gec,Brdar:2020quo}. Within the framework of the minimally extended SM with right-handed SU(2) gauge singlets, the diagonal magnetic moments for massive Dirac neutrinos are evaluated to be~\cite{Giunti:2024gec}
\begin{equation}
\mu_{\nu}\simeq 3.2 \times 10^{-19}\left(\frac{m_i}{1\rm eV}\right)\mu_{\rm B}\,,
    \label{mu_nuD}
\end{equation}
where $m_i$ is the mass of the \(i\)-th mass eigenstate and \(\mu_{\rm B}\) stands for the Bohr magneton. For the ultrarelativistic neutrinos, we can approximate their momenta $p$ with $p \gg m_i$ and $p \gg \mu_\nu B_{\perp}$, where $B_{\perp}$ is the strength of the transverse magnetic field. In~\cite{Alok:2024xeg}, the coherence of neutrino systems was investigated within the resource theoretic framework. The study demonstrated that, for SFOs, the coherence scale can be significantly extended to astrophysical distances. This extension could potentially enable future quantum technologies at such scales, including communication using neutrinos both within and beyond the Milky Way. 

The Dirac equation for the massive Dirac neutrino spin eigenstate $|\nu_{i}^{s}\rangle$ propagating in the presence of an arbitrarily oriented magnetic field $\boldsymbol{B}$ can be given as,
\begin{equation}
     \left[\gamma_{\mu}p^{\mu} - m_{i} - \mu_{i}\boldsymbol{\left(\Sigma\cdot B\right)} \right] \nu_{i}^{s}(p) = 0,
     \label{evol}
\end{equation}
where $\mu_i$ stands for the diagonal Dirac neutrino magnetic moment, $\boldsymbol{\Sigma}$ is given by the Pauli matrices $\sigma_i$ (generators of $\rm SU(2)$ group) as $\Sigma_i=\begin{pmatrix}
    \sigma_i&&0\\0&&\sigma_i
\end{pmatrix}$ and the Hamiltonian ($\hat{H}_{i}$) representing the system here can be obtained simply by hitting Eq. \eqref{evol} with $\gamma_{0}$ as,
\begin{equation}
    \hat{H}_{i} = \gamma_{0} \boldsymbol{\gamma \cdot p} + \mu_{i}\gamma_{0} \boldsymbol{\left(\Sigma\cdot B\right)} + m_{i} \gamma_{0}\,.
    \label{hamiltonian}
\end{equation}
In Eq. \eqref{evol}, $s\, (= \pm 1)$ are the eigenvalues of the spin operator $\hat{S}_{i}$ such that $\left[\hat{H}_{i},\hat{S}_{i}\right]=0$,
\begin{equation}
    \hat{S}_{i} = \frac{m_{i}}{\sqrt{m_{i}^{2}\boldsymbol{B}^{2} + \boldsymbol{p}^{2}\boldsymbol{B}_{\perp}^{2}}} 
    \left[ \boldsymbol{\left(\Sigma\cdot B\right)} - \frac{i}{m_i} \gamma_{0} \gamma_{5} [\boldsymbol{\Sigma \times \boldsymbol{p}}] \boldsymbol{\cdot} \boldsymbol{B} \right]\,.
    \label{spin-op}
\end{equation}

Here the neutrinos are assumed to be propagating in the $z$-direction, thus their momentum is $\boldsymbol{p} = p_{z}$ and the  magnetic field is given by, $\boldsymbol{B} = (B_{\perp}, 0, B_{\parallel})$. The energy spectrum of neutrino can be expressed as \cite{Popov:2019nkr},
\begin{equation}
    E_{i}^{s} = \left(m_{i}^2 + p^2 + \mu_{i}^{2} \boldsymbol{B}^{2} + 2\mu_{i}s\sqrt{m_{i}^{2}\boldsymbol{B}^{2} + \boldsymbol{p}^{2}{B}_{\perp}^{2}}\right)^{\frac{1}{2}}\,.
    \label{en-gen}
\end{equation} 

For ultra relativistic neutrinos, we can approximate their momenta $p \gg m_i$ and $p \gg \mu_\nu B$, also within the considered environment for neutrino propagation, given the smallness of the neutrino magnetic moment and the upper bounds on the magnetic fields for a given neutrino mass species, we can assume $\mu_i B\ll m_i$. Thus, the energy for the neutrinos can be approximated in this limit as,
\begin{equation}
    E_{i}^{s} \approx p + \frac{m_{i}^{2}}{2p} + \mu_{i}sB_{\perp}. 
    \label{energyf}
\end{equation}
Under the assumption that all the neutrino states possess an equal magnitude of the magnetic moment, the oscillation phase of SFOs is given by the frequency $\xi_{ji}^{s's}$ that can be denoted as
\begin{equation}
    \xi_{ji}^{s's}=E_{j}^{s'}-E_{i}^{s} = \frac{\Delta m_{ji}^{2}}{2E} + \mu_{\nu}(s'-s)B_\perp\;,
    \label{freq}
\end{equation}
where $\Delta m_{ji}^{2}=\Delta m_{j}^{2}-\Delta m_{i}^{2}$, $E$ is the neutrino energy in the massless limit, and $s$ and $s'$ are the helicity indices.

The mixing between the flavor and mass eigenstates is described by Eq.~\eqref{3mix}.
Since the helicity mass eigenstates are not stationary in the presence of an external magnetic field, they must be decomposed into the stationary spin mass eigenstates, as given by
\begin{equation}
    \ket{\nu_i ^h (t)}=\sum_{s=-1}^{+1} k_{i}^s  \ket{\nu_i ^s (t)}\;.
    \label{tres}
\end{equation}

Now by applying Eq.~\eqref{tres} in Eq.~\eqref{3mix}, we arrive at
\begin{align}
    \ket{\nu_\alpha ^h (t)} &= \sum_{i=1}^{3} \sum_{s=-1}^{+1} U_{\alpha i}^{\ast} k_{i}^{s} \ket{\nu_i ^s (t)} \,\label{cinco} \\ 
    &= \sum_{i=1}^{3} \sum_{s=-1}^{+1} U_{\alpha i}^{\ast} k_{i}^{s} e^{-i E_{i}^s t} \ket{\nu_i ^s (0)} \,.
    \label{seis}
\end{align}
Subsequently, we can express the initial stationary spin mass eigenstates in terms of the initial helicity flavor eigenstates as follows:
\begin{equation}
    \ket{\nu_i ^s }= \sum_{\beta=e}^{\tau} \sum_{s=-1}^{+1} U_{\beta i} k_{i}^{s \ast} \ket{\nu_\beta ^{h'}}\;.
    \label{ocho}
\end{equation}
For the three-flavor mixing case, if we consider left-handed Dirac neutrinos of the initial flavor $\alpha$, the evolution of the flavor eigenstate $|\nu_{\alpha}^{L}(l)\rangle$ undergoing SFOs can be expressed as 
\begin{equation}
    \ket{\nu_\alpha ^L (l)}=\sum_{h'=L}^{R}\sum_{\beta=e}^{\tau} \sum_{i=1}^{3} \sum_{s=-1}^{+1} U_{\alpha i}^{\ast}U_{\beta i} C_{is}^{Lh'}e^{-i E_i^s l}\ket{\nu_\beta ^{h'}}\,,
\label{nueve}
\end{equation}
where $C_{is}^{hh'}=\langle \nu_i^{h'} | \hat{P}_i^s | \nu_i^h \rangle$ is the plane-wave expansion coefficient which can be calculated by matrix elements of the projection operator \(\hat{P}_i^s = \hat{P}_i^{\pm} \equiv |\nu_i^{\pm} \rangle \langle \nu_i^{\pm}|\). Note that in the ultrarelativistic limit ($p\gg m_i$), we disregard terms of the order $\mathcal{O}\left({m_i/p}\right)^2$ and higher. This simplification reduces the plane-wave expansion to $C_{is}^{LL}\simeq {1}/{2}$ and $C_{is}^{LR}\simeq-{s}/{2}$. In the above expression, by factoring out the spin and helicity summations, we can restore the evolution of the flavor state $|\nu_{\alpha} (l)\rangle$ for FOs as 
\begin{eqnarray}
\ket{\nu_{\alpha} (l)} &=& \sum_{\beta=e}^{\tau} \sum_{i=1}^{3} U_{\alpha i}^{\ast}U_{\beta i} e^{-i E_i l}\ket{\nu_\beta} \,.
  \label{evol-3fo}
\end{eqnarray} 

Considering the simplified case with two flavors  of $\alpha=(e,\mu)$, the evolution of the flavor eigenstate $|\nu_{e,\mu}^L\rangle$ can be expressed in the spin-flavor basis as
\begin{eqnarray}  
\label{evol-sfo}
\ket{\nu_{e,\mu}^L (\theta,l)} =  f_1(\theta,l) \ket{\nu_{e,\mu}^L} + f_2(\theta,l) \ket{\nu_{e,\mu}^R} + f_3(\theta,l) \ket{\nu_{\mu,e}^L} +f_4(\theta,l)\ket{\nu_{\mu,e}^R} \,,
\end{eqnarray}
where the coefficients are given by,
\begin{widetext}
\begin{eqnarray}
2f_1(\theta,l) &=& 
    \cos^2 \theta e^{-i E_i^+ l} + \sin^2\theta e^{-i E_j^+ l} 
    + \cos^2\theta  e^{-i E_i^- l}  + \sin^2\theta  e^{-i E_j^- l} \nonumber\\
2f_2(\theta,l) &=& 
    \cos^2\theta  e^{-i E_i^- l} + \sin^2\theta  e^{-i E_j^- l} - \cos^2\theta e^{-i E_i^+ l} - \sin^2\theta  e^{-i E_j^+ l} \nonumber\\
2f_3(\theta,l) &=&      -  \sin \theta\cos\theta\bigg[
    e^{-i E_i^+ l} - e^{-i E_j^+ l} + e^{-i E_i^- l} - e^{-i E_j^- l}
    \bigg] \nonumber\\
2f_4(\theta,l) &=&     
    -  \sin\theta\cos\theta \bigg[
     e^{-i E_i^- l} - e^{-i E_j^- l} - e^{-i E_i^+ l} + e^{-i E_j^+ l} \bigg]\,
\end{eqnarray}
\end{widetext}
with $\theta$ being the flavor mixing angle.
From the state evolution, we can deduce the $4\times4$ density matrix $\rho_{e,\mu}^{\rm SFO} (\theta,l)=|\nu_{e,\mu}^{L}(\theta,l)\rangle\langle\nu_{e,\mu}^{L}(\theta,l)|$. In this matrix, the diagonal elements correspond to the probabilities of the two-flavor SFOs.

The evolution of standard two-flavor FOs can be derived by setting the spin eigenvalues equal, i.e., $s = s'$, and eliminating the spin-flipping degrees of freedom in Eq.~\eqref{evol-sfo}. Consequently, the second and last terms in the evolution equation of SFOs vanish. Additionally, the first and third terms simplify to $f'_1(\theta,l)$ and $f'_3(\theta,l)$, which are given by $(\cos^2 \theta e^{-iE_i l} + \sin^2 \theta e^{-iE_j l})$ and $\sin \theta \cos \theta\left(e^{-iE_j l} - e^{-iE_i l}\right)$, respectively. 
This simplification leads to a reduced $2\times 2$ density matrix, $\rho_{e,\mu}^{\rm FO} (\theta,l)=|\nu_{e,\mu}(\theta,l)\rangle\langle\nu_{e,\mu}(\theta,l)|$. It is also evident from Eq.~\eqref{freq} that by equating the spin eigenvalues, we obtain the phase for FOs, which can then be used to calculate the elements of the aforementioned density matrix.

In the context of three-flavor oscillations (3FOs) of neutrinos, the complex phase $\delta_{\rm CP}$ appears in the PMNS mixing matrix and is anticipated to contribute to the imaginarity measure of the system. Imaginarity measures remain nontrivial even in systems with a zero $\delta_{\rm CP}$, and in scenarios of two-flavor mixing which is described by a simple rotation matrix parameterized by the mixing angle $\theta$ between the two flavors. Therefore, it is essential to examine the imaginarity in the context of two-flavor mixing for both standard FOs and SFOs, as well as in 3FOs, taking into account the dependence on $\delta_{\rm CP}$.

\section{Imaginarity in Neutrino Systems}

We begin by quantifying the imaginarity in the coherent superposition of neutrino states in the systems previously defined, using the imaginarity measure based on the $\ell_1$-norm. For the two-flavor case involving FOs, the $\ell_1$-norm of imaginarity can be calculated using the density matrices deduced in the previous section. It is important to note that the imaginary components appear only in the off-diagonal elements of the density matrix. Therefore, only those elements should be considered when computing the $\ell_1$-norm of imaginarity. The $\ell_1$-norm of imaginarity in 2FOs framework is calculated to be,

\begin{equation}
\mathcal{I}_{\ell_1}\left(\rho_{e,\mu}^{\rm FO}\right)=\sum_{i \neq j} |\mathrm{Im}\left(\rho_{e,\mu}^{\rm FO}\right)_{ij}|=\left|\sin 2\theta\sin{\left(\frac{\Delta m_{ji}^{2}}{2E}l\right)}\right|\,.
\label{iml1-1}
\end{equation}
 It is important to note that, despite the initial neutrino states being different, the $\ell_1$-norm of imaginarity turns out to be the same for them. We can also immediately observe that the $\ell_1$-norm of imaginarity is nonzero even for two-flavor FOs. This finding underscores the significance of imaginarity as a resource coming from the intrinsic propagation dynamics of neutrinos.
Next, to calculate the relative entropy of imaginarity in two-flavor FOs, we evaluate $|\langle\nu_{e,\mu}^{\ast}(\theta,l)|\nu_{e,\mu}(\theta,l)\rangle|$ as follows:

\begin{widetext}
\begin{equation}
|\langle\nu_{e,\mu}^{\ast}(\theta,l)|\nu_{e,\mu}(\theta,l)\rangle|=\\ \sqrt{\cos^4 \theta+\sin^4 \theta+2\sin^2 \theta \cos^2 \theta \cos{\left(\frac{\Delta m_{ji}^{2}}{E}l\right)}}\,.
    \label{relfo}
\end{equation}

\end{widetext}

From the above expression, we can derive the measure $\mathcal{I}_{r}\left(|\nu_{e,\mu}(\theta,l)\rangle\right)$ by applying Eq.~\eqref{relfo} in Eq.~\eqref{rel}. Although one might naively expect that imaginarity as a resource could only arise from complex phases in the mixing matrix, it turns out that time evolution alone can lead to nonzero imaginarity, even in the absence of a $CP$-violating phase. This suggests that imaginarity can still play a role as a resource under such conditions. We have demonstrated this point by using the two proper measures of imaginarity.
The imaginarity metric based on the $\ell_1$-norm, as shown in Eq.~\eqref{iml1-1}, captures the imaginarity present in the intrinsic coherent superposition within the two-flavor FOs. In contrast, the metric presented in Eq.~\eqref{relfo} quantifies the amount of imaginarity as a resource by measuring the entropic distance between the state \(\rho\) and its real part \(\text{Re}(\rho)\).

Now, we are going to quantify the imaginarity in SFOs between two flavors. The additional spin-flipping degrees of freedom arise because neutrinos possess a magnetic moment, which may interact with the pervasive magnetic field environment. Moreover, similar to the $\ell_1$-norm of coherence \cite{Song:2018bma, Alok:2024xeg}, the $\ell_1$-norm of imaginarity is also a basis-dependent measure. Therefore, it is important to examine its behavior in systems with additional degrees of freedom. In the context of neutrinos, spin-flavour oscillations provide such a system for investigation.

By utilizing the $4\times 4$ density matrix $\rho_{e,\mu}^{\rm SFO} (\theta,l)=|\nu_{e,\mu}^{L}(\theta,l)\rangle\langle\nu_{e,\mu}^{L}(\theta,l)|$ derived from the spin-flavor evolution in Eq.~\eqref{evol-sfo}, the $\ell_1$-norm of imaginarity can be computed as follows:

\begin{widetext}
\begin{align}
\mathcal{I}_{\ell_1}\!\left(\rho_{e,\mu}^{\rm SFO}\right)
&= |\sin\theta \cos\theta|
\Bigg\{
\Big|2(\cos^2\theta-\sin^2\theta)\sin\!\left(\xi_{ii}^{+-} l\right)
\cos\!\left(\frac{\Delta m_{21}^{2}}{2E}l\right)\Big|
\nonumber\\
&\quad
+ \Big|\sin\!\left(\frac{\Delta m_{21}^{2}}{2E}l\right)\Big|
\Big(|\cos\!\left(\xi_{ii}^{+-}l\right)-1|
+|\cos\!\left(\xi_{ii}^{+-}l\right)+1|\Big)
\Bigg\}.
\label{l1sfo}
\end{align}

The relative entropy of imaginarity $\mathcal{I}_{r}\!\left(|\nu_{e,\mu}(\theta,l)\rangle\right)$ is calculated via Eq.~\eqref{rel} using
the overlap $|\langle \psi^*|\psi\rangle|$,

\begin{align}
|\langle\nu_{e,\mu}^{L*}(\theta,l)|\nu_{e,\mu}^{L}(\theta,l)\rangle|
&=
\Bigg[
\frac{(\cos^4\theta+\sin^4\theta)}{2}\Big(1+\cos\!\left(2\xi_{ii}^{+-}l\right)\Big)
\nonumber\\
&\quad
+\frac{\sin^2\theta \cos^2\theta}{2}
\Bigg\{\cos\!\left(2\xi_{ji}^{+-}l\right)+\cos\!\left(2\xi_{ji}^{-+}l\right)
+2\cos\!\left(\frac{\Delta m_{ij}^{2}}{E}l\right)\Bigg\}
\Bigg]^{1/2}.
\label{relsfo}
\end{align}
\normalsize
\end{widetext}
Note that $\xi_{ji}^{\pm\mp}$ which appears in both of the above expressions in Eqs.~\eqref{l1sfo} and ~\eqref{relsfo}, represents the the SFO frequency defined in Eq.~\eqref{freq}. 
It should also be noted that both measures of the SFO system, as expressed in the equations above, will reduce to those of the FO system in the absence of spin-flipping caused by the interaction of neutrinos with an external magnetic field, due to their nonzero magnetic moment. This demonstrates that neutrino systems inherently possess a component of imaginarity, making them viable as a resource for quantum information tasks, even if the complex phase $\delta_{\rm CP}$ in the leptonic sector is found to be zero.

In order to analyse the quantification of imaginarity in neutrino systems, we examine the behaviour of the $\ell_1$-norm of imaginarity in the short-baseline reactor experiment Daya Bay~\cite{DayaBay:2012fng} and the medium-baseline experiment 
KamLAND~\cite{KamLAND:2002uet}. Nuclear reactors serve as intense sources of coherent antineutrino fluxes and  offers critical insights into neutrino oscillation parameters. Daya Bay operates at 1.5 km baseline and detects reactor antineutrinos in the energy range E = 1–10 MeV. KamLAND utilizes a baseline of $L$ = 180 km, over same antineutrino energy spectrum. We focus on these experiments because our analysis is restricted to neutrinos oscillating in vacuum. 

For Daya Bay, the matter effects are negligible due to the small baseline, and the oscillations are considered within the standard vacuum oscillation framework~\cite{DayaBay:2018yms}. In KamLAND, the matter effects are non-zero but small~\cite{KamLAND:2008dgz}. In two-flavour oscillations, the oscillation parameters can be directly replaced by the effective oscillation parameters in constant matter density. Since the imaginarity measures depend on these parameters, such a modification would not lead to any change in the behaviour of the measures, as the functional form remains the same. To highlight only the dynamical behaviour, we restrict ourselves to the vacuum approximation. In long-baseline experiments, matter effects would be large, which is important to take into consideration in future possible works.

\begin{figure}[H]
    \centering
    \includegraphics[width=0.47\linewidth]{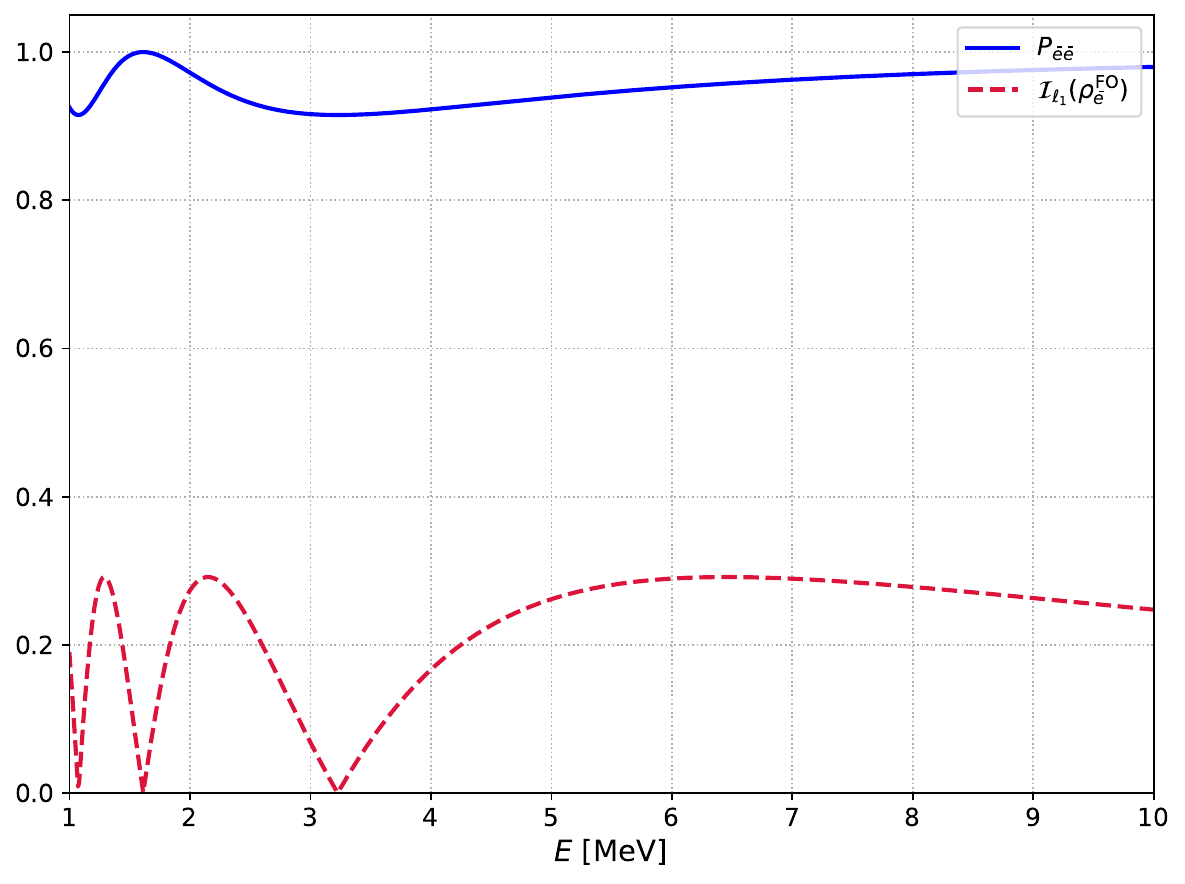}
    \includegraphics[width=0.47\linewidth]{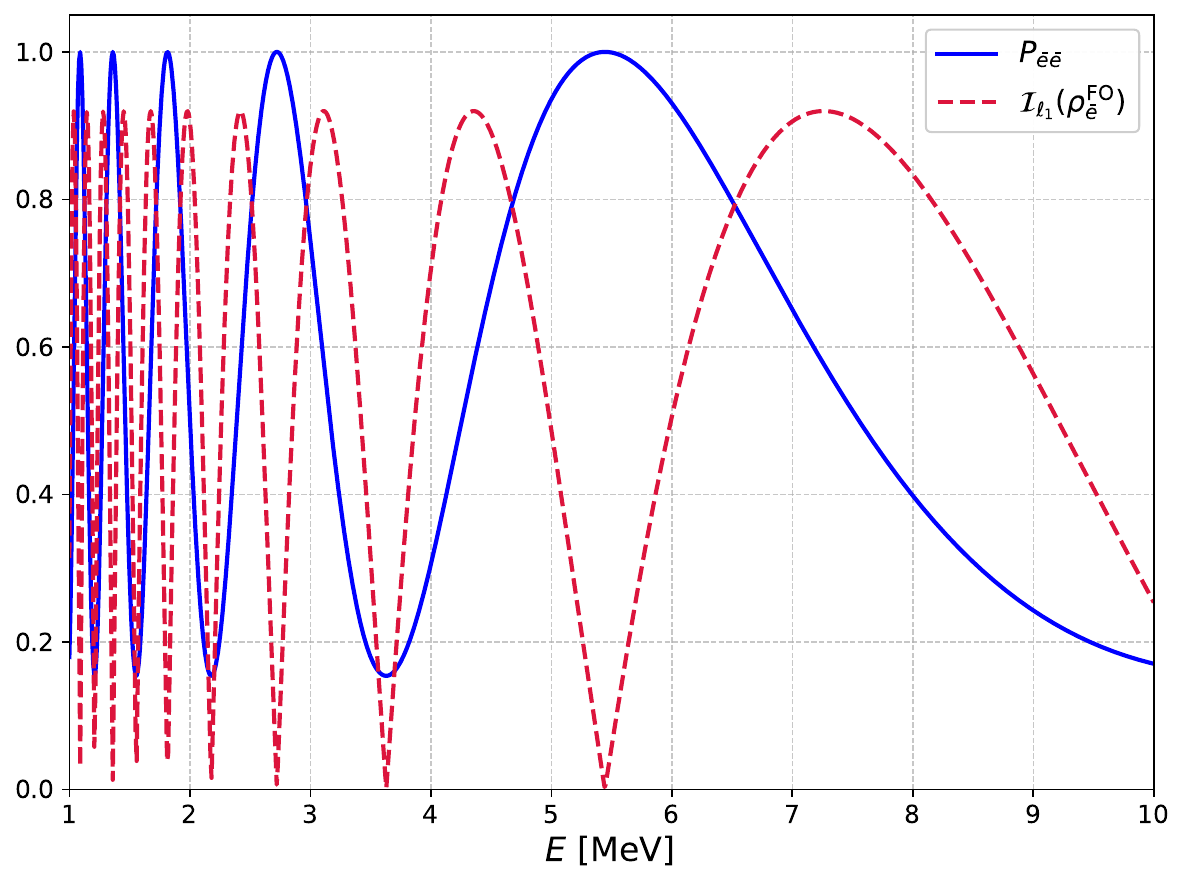}
    \caption{The electron antineutrino survival probability (blue) and the $\ell_1$-norm of imaginarity (red) are shown as functions of antineutrino energy for (a) the Daya Bay experiment and (b) the KamLAND experiment. We have considered the best fit values of oscillation parameters from Ref.~\cite{Esteban:2024eli}.}
    \label{fig:reactors}
\end{figure}

The two plots in Fig.~\ref{fig:reactors} shows how the $\ell_1$-norm of imaginarity, i.e. $\mathcal{I}_{\ell_1}^{\text{FO}}$, behaves in reactor neutrino experiments for different baselines. The amplitude of the imaginarity curve is determined by the mixing angles. In the case of KamLAND, the relevant mixing angle is $\theta_{12}$, which is relatively large, whereas Daya Bay is primarily sensitive to $\theta_{13}$, which is significantly smaller. Hence, the magnitude of the imaginarity is larger for KamLAND than for Daya Bay. This illustrates that the overall strength of imaginarity in the neutrino state is governed by the mixing angle, while the locations of its maxima and zeros are controlled by the oscillation phase accumulated during propagation depending on the distinct mass-squared difference in both cases. It should also be noted that the solar mixing parameters relevant for KamLAND, as well as $\theta_{13}$ and the atmospheric mass splitting relevant for Daya Bay, are now measured with high precision (typically at the sub-percent to percent level). Therefore, varying these parameters within their experimental uncertainties does not alter the qualitative behavior of the imaginarity measures. If we compare the behaviour of $\ell_1$-norm of imaginarity with the oscillation probability, we see that, $\mathcal{I}_{\ell_1}^{\text{FO}}$ is maximum where the survival probability $P_{\bar{e}\bar{e}}$ is changing most rapidly, reflecting maximal quantum interference between flavor states. In contrast, $\mathcal{I}_{\ell_1}^{\text{FO}}$ vanishes when $P_{\bar{e}\bar{e}}$ reaches a maximum or minimum where the neutrino state actually realigns with a particular flavor eigenstate i.e. it becomes stationary, and exhibits no mixing behaviour.

\begin{figure}[H]
    \centering   
    \includegraphics[width=0.48\linewidth]{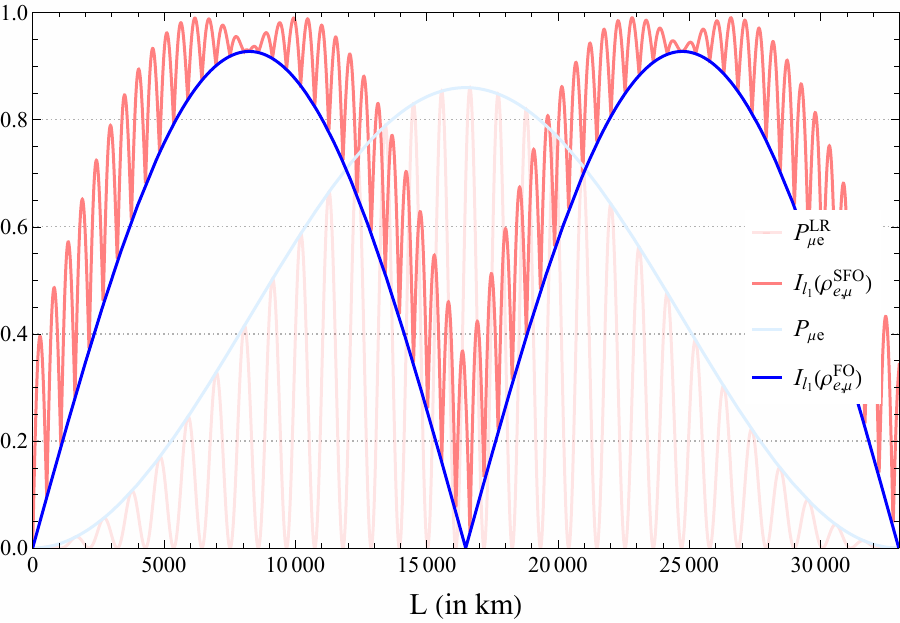} \hspace{1em}
    \includegraphics[width=0.48\linewidth]{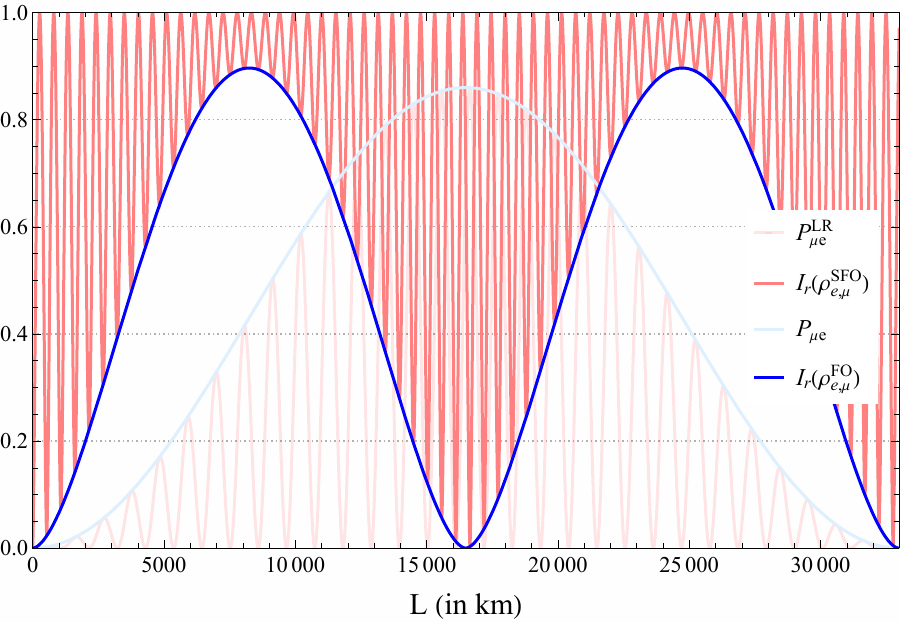}
    \caption{The left panel shows the $\ell_1$-norm of imaginarity, while the right panel displays the relative entropy of imaginarity, both as functions of the propagation length (in km) for high-energy GeV muon neutrinos. The two-flavor FO system is represented in blue, and the two-flavor SFO system is in red, with the flavor transition probability shown in light blue and the spin-flavor transition probability in light red using solar oscillation parameters~\cite{Esteban:2024eli}.}
    \label{fig:plot}
\end{figure}

To provide a quantitative illustration of the properties of the imaginarity measures across different neutrino quantum systems, Fig.~\ref{fig:plot} displays the $\ell_1$-norm, $\mathcal{I}_{\ell_1}$,  (left panel) and the relative entropy $\mathcal{I}_{r}$ (right panel) of imaginarity as functions of the propagation length for high-energy GeV muon neutrinos. For comparison, the oscillation probabilities of both FOs and SFOs systems are also presented. 
The figure demonstrates a similar behaviour to that of the reactor neutrino case, that imaginarity vanishes at the extremum points of flavor oscillation probabilities. It is also evident that, in the system of FOs, the imaginary component reaches its maximum when the transition and survival probabilities are comparable, around 0.5, corresponding to a coherent superposition of flavor states with nearly equal amplitudes. In contrast, imaginarity vanishes at the extrema of the oscillation probability, where the imaginary parts of the off-diagonal density matrix elements disappear in the case of $\ell_1$-norm of imaginarity.

In Fig.~\ref{fig:plot}, we also illustrate the nature of imaginarity in the system of SFOs, as compared to the standard FO system. For this illustration, we consider a magnetic field of approximately $10^{8}$ G and a neutrino magnetic moment of $\mu_\nu \sim 10^{-12} \mu_{\rm B}$, which is close to the current experimental upper limit~\cite{Giunti:2024gec}. These values of parameters are chosen to allow a direct comparison of the behavior of imaginarity in the FO and SFO systems. Although the spin-precession phase introduces an additional modulation to the imaginarity curve in the SFO system, the positions of its maxima and zeros remain aligned with those determined by the FO phase.

\begin{figure}[H]
    \centering   
    \includegraphics[width=0.48\linewidth]{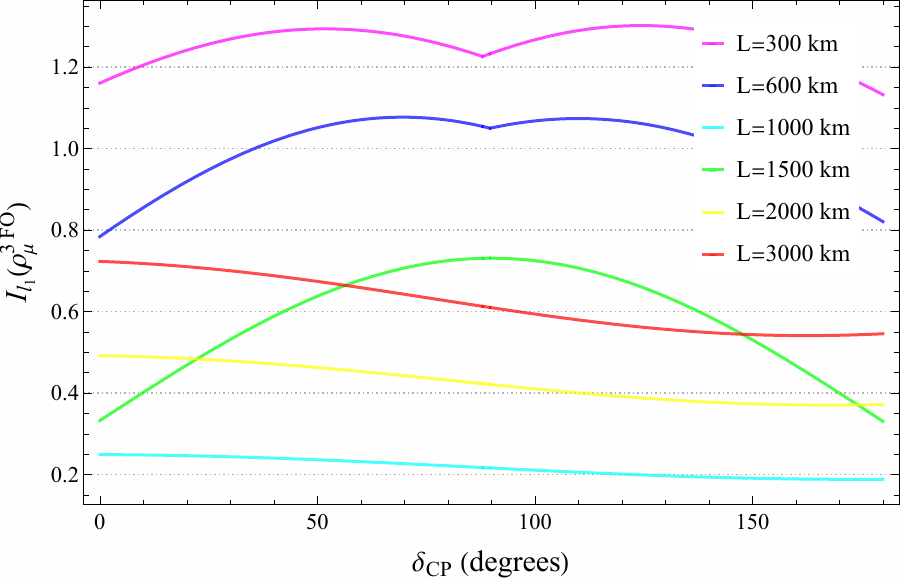} \hspace{1em}
    \includegraphics[width=0.48\linewidth]{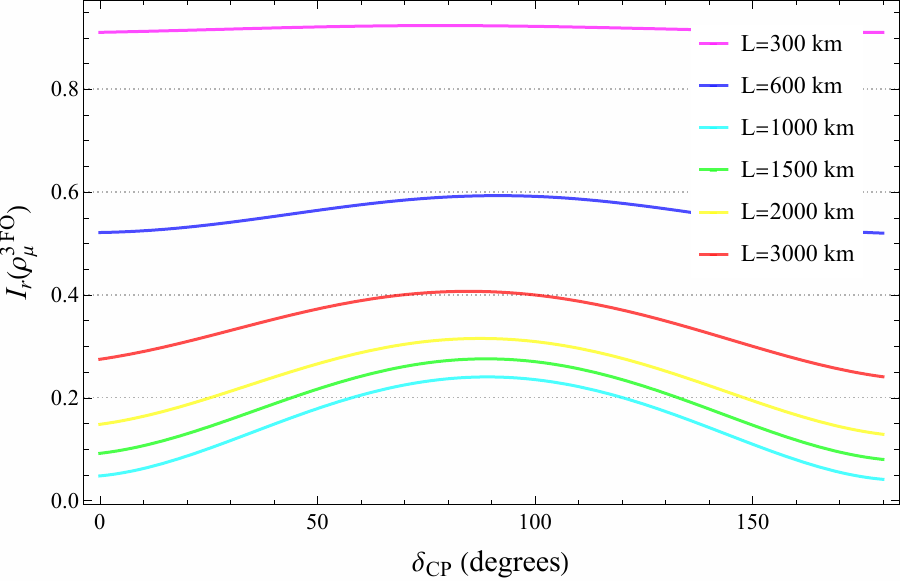}
    \caption{The figure on the left panel illustrates the  $\ell_1$-norm of imaginarity, while the right panel shows the relative entropy of imaginarity, both plotted as a function of the $CP$-violating phase \( \delta_{\rm CP} \), for different propagation lengths (in km) of high-energy GeV muon neutrinos.}
    \label{fig:plot2}
\end{figure}
Finally, it will be intriguing to observe how the presence of a $CP$-violating phase affects imaginarity. To investigate this, we examine the \( \ell_1 \)-norm of imaginarity for 3FOs of neutrinos. In this case, we consider the \( 3\times 3 \) density matrix defined by the three-flavor evolution in Eq.~\eqref{evol-3fo}, where the matrix elements will depend on \( \delta_{\rm CP} \). The results, shown in Fig.~\ref{fig:plot2}, indicate that, similar to the case of two-flavor FOs, the imaginarity measures remain non-zero even when \( \delta_{\rm CP}=0 \). This reaffirms that imaginarity can be quantified from the dynamics of the neutrino system and persists even if the value of the $CP$-violating phase in the leptonic sector is zero.  Furthermore, the figures demonstrate that for a given \( L \), depending on the non-zero value of \( \delta_{\rm CP}\), both the \( \ell_1 \) norm and the relative entropy of imaginarity can be enhanced or suppressed compared to the case of \( \delta_{\rm CP}=0 \). For certain values of \( L \), the dependence of imaginarity on \( \delta_{\rm CP}\) is minor, while for others, the deviation from that of \( \delta_{\rm CP}=0 \) could be significant. This behaviour also indicates the oscillation parameter dependence of imaginarity measures, since their magnitude is governed by the baseline, energy and mixing parameters. 

The $\ell_1$-norm of coherence quantifies the ``quantumness" of a quantum state by accounting for the off-diagonal elements of its density matrix, which represent the state's superposition in a given basis. In contrast, the $\ell_1$-norm of imaginarity specifically captures the imaginary parts of these off-diagonal elements. While both measures are basis-dependent and vanish for purely diagonal states, they highlight different facets of quantum behavior. We show that the phase accumulated during the propagation (or time evolution) of neutrino states that generates imaginarity in the system, can also be interpreted as a quantum resource in neutrino oscillations. Hence, imaginarity as a resource does not require $CP$ violation as the only origin.


\section{Conclusions}
In this study, we have explored the quantification of imaginarity in neutrino flavor and spin-flavor oscillations using a resource-theoretic framework, which addresses the fundamental question of why quantum mechanics employs complex numbers instead of real ones. This framework provides multiple quantifiers to capture different aspects of the inherent quantumness in a system. We have focused on two such quantifiers: the $\ell_1$-norm of imaginarity, which reflects imaginarity in coherent superpositions, and the relative entropy of imaginarity, which quantifies imaginarity as a resource by measuring the entropic distance between a state $\rho$ and its real part $\text{Re}(\rho)$.

In neutrino systems, the complex $CP$-violating phase arises naturally within the three-flavor mixing framework. Importantly,  we have demonstrateed that imaginarity persists as a nontrivial feature even in the two-flavor approximation, relevant to both flavor and spin-flavor oscillations.
The measures of imaginarity reaches their maximum where the probabilities \textit{change} most rapidly, reflecting maximal quantum interference between flavor states. Conversely, imaginarity vanishes when probabilities attains either a maximum or minimum, indicating that the neutrino state becomes aligned with a pure flavor eigenstate and exhibits no mixing dynamics.

This study also indicates that imaginarity as a resource in neutrino systems is not exclusively dependent on the $CP$-violating phase. More importantly, the imaginarity can arise in the neutrino system as a resource, from the intrinsic quantum dynamics of the neutrino evolution itself. Neutrinos, being naturally oscillating quantum systems, offer an ideal platform to explore this quantum nature. Furthermore, if the imaginarity measures are experimentally observed to be nonzero, this would serve as direct evidence for the fundamental importance of complex numbers in quantum theory, revealed uniquely through neutrino physics.

\section{Acknowledgements}
The current work is dedicated to Prof.~Ashutosh Kumar Alok, who unexpectedly passed away on 10 October 2024, during the final stage of the manuscript preparation. T.J. Chall and N.R.S. Chundawat are grateful to Prof.~Ashutosh Kumar Alok for his invaluable guidance and support throughout their academic endeavors. The work of Y.F. Li is partially supported by National Natural Science Foundation of China under Grant Nos.~12075255 and 11835013.

\end{document}